# Direct observation of layer-stacking and oriented wrinkles in multilayer hexagonal boron nitride

Lingxiu Chen[1,2,3], Kenan Elibol[4,5], Haifang Cai[6], Chengxin Jiang[1,7], Wenhao Shi[6], Chen Chen[1,2], Hui Shan Wang[1,2], Xiujun Wang[1,2], Xiaojing Mu[8], Chen Li[9], Kenji Watanabe[10], Takashi Taniguchi[10], Yufeng Guo[6]*, Jannik C. Meyer[4,11]*, Haomin Wang[1,2]*

[1] State Key Laboratory of Functional Materials for Informatics, Shanghai Institute of Microsystem and Information Technology, Chinese Academy of Sciences, Shanghai 200050, China
[2] Center of Materials Science and Optoelectronics Engineering, University of Chinese Academy of Sciences, Beijing 100049, China
[3] School of Materials Science and Physics, China University of Mining and Technology, Xuzhou 221116, China
[4] Faculty of Physics, University of Vienna, Vienna 1090, Austria
[5] School of Chemistry, Trinity College Dublin, Dublin 2, Ireland
[6] State Key Laboratory of Mechanics and Control of Mechanical Structures and MOE Key Laboratory for Intelligent Nano Materials and Devices, College of Aerospace Engineering, Nanjing University of Aeronautics and Astronautics, Nanjing 210016, China
[7] School of Physical Science and Technology, ShanghaiTech University, Shanghai 201210, China
[8] Key Laboratory of Optoelectronic Technology & Systems Ministry of Education International R & D Center of Micro-Nano Systems and New Materials Technology, Chongqing University, Chongqing 400044, China
[9] Electron Microscopy for Materials Research (EMAT), University of Antwerp, Antwerp 2020, Belgium
[10] National Institute for Materials Science, Tsukuba 305-0044, Japan
[11] Institute for Applied Physics and Natural and Medical Sciences Institute, University of Tübingen, Tübingen 72076, Germany

E-mail: hmwang@mail.sim.ac.cn, yfguo@nuaa.edu.cn and jannik.meyer@uni-tuebingen.de

## Abstract

Hexagonal boron nitride ($h$-BN) has long been recognized as an ideal substrate for electronic devices due to its dangling-bond-free surface, insulating nature and thermal/chemical stability. These properties of the $h$-BN multilayer are mainly determined by its lattice structure. Therefore, to analyse the lattice structure and orientation of $h$-BN crystals becomes important. Here, the stacking order and wrinkles of $h$-BN are investigated by transmission electron microscopy (TEM). It is experimentally confirmed that the layers in the $h$-BN flakes are arranged in the AA′ stacking. The wrinkles in a form of threefold network throughout the $h$-BN crystal are oriented along the armchair direction, and their formation mechanism was further explored by molecular dynamics simulations. Our findings provide a deep insight about the microstructure of $h$-BN and shed light on the structural design/electronic modulations of two-dimensional crystals.

## 1. Introduction

Hexagonal boron nitride ($h$-BN) is an insulating layered material that has attracted intense attention due to its wide bandgap and extreme stability [1-4]. Single-layer $h$-BN is covalently bonded through sp$^2$ hybridization and forms multilayer $h$-BN via van der Waals forces [5]. Multilayer $h$-BN has been widely used as either a substrate or an encapsulating layer to produce stable van der Waals heterostructures composed of various two-dimensional crystals [3,6-11]. Theoretical studies showed that the most stable stacking order in multilayer $h$-BN is AA′ [12], which is quite different from the stacking order in graphite [5,13-16]. The aligned porous stacking order of $h$-BN plays an important role in the ability of multilayer $h$-BN to isolate hydrogen by a plasma treatment, while due to its different stacking order, such isolation cannot be performed using graphite [17]. In layered materials, the stacking order may greatly influence structural deformations that have been extensively studied in mechanics and optics [18,19]. Additionally, deformations such as wrinkles and ripples can strongly influence the physical properties of materials [18,20-24]. Moreover, wrinkles can form potential edgeless nanocapillaries for the transport of ions and molecules [25-27]. *h*-BN films supported by other substrates may form wrinkles due to the negative thermal expansion coefficient of *h*-BN [19,28-30]. Detailed investigations of the structure of *h*-BN wrinkles are still necessary for potential application of *h*-BN.

Here, we investigated the stacking order and wrinkles of multilayer *h*-BN by conventional and scanning transmission electron microscopy (TEM/STEM). Furthermore, high-temperature annealing in oxygen was carried out to investigate the anisotropic etching of the *h*-BN multilayer. Atomic force microscopy (AFM) measurements showed that the triangle-shaped holes etched in the adjacent layers of multilayer *h*-BN have opposite orientations. This provides experimental confirmation of the AA′ stacking in the multilayer *h*-BN. Molecular dynamics simulations were performed to understand the formation mechanism of oriented wrinkles on multilayer *h*-BN. The presence of wrinkles provides an easy and rapid method for the identification of the crystallographic orientation of *h*-BN.

## 2. Results

To investigate the lattice structures of multilayer *h*-BN, *h*-BN flakes prepared by exfoliation were examined by TEM and STEM. **Figure 1a** shows a plan-view STEM-MAADF image of a multilayer *h*-BN flake. The e-beam energy is set at 60 keV to minimize the damage to the lattice of *h*-BN. The flake has a thickness of 5 atomic layers. The periodic bright spots correspond to individual atomic columns in the honeycomb arrangement with a lattice constant of 2.5 Å, which is in good agreement with the value recorded in the literature (**Figures 1d-e**) [5,31]. As shown in **Figure 1a**, the *h*-BN multilayer exhibits a porous structure. It is observed that the pores of the hexagonal network in each layer are precisely aligned. We also find that each atomic column along the direction of the incident electron beam shows similar contrast. It is found that the STEM image shown in Figure 1a matches very well with the simulated result, which is exhibited in Figure S2. Thus, the similar contrast confirms that the *h*-BN multilayer is found in the AA′ stacking arrangement [32,33].

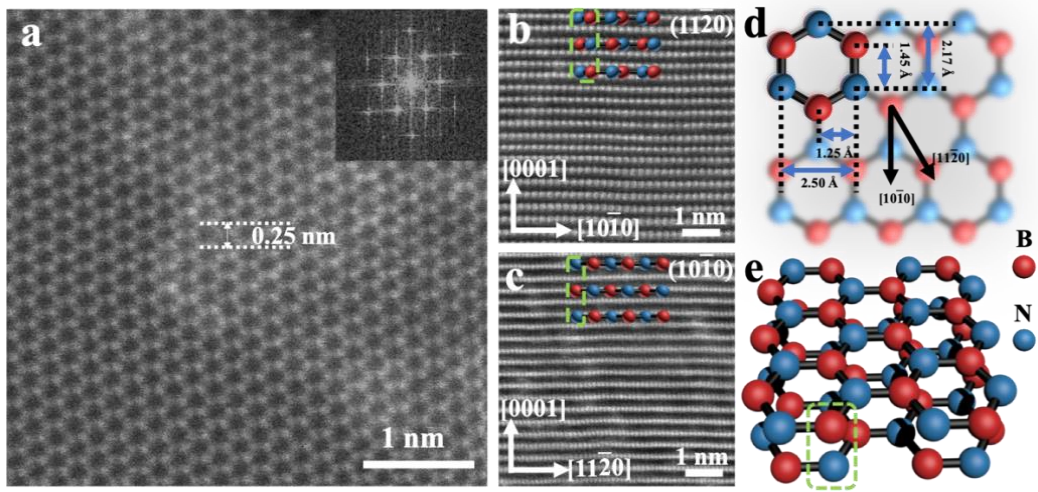

**Figure 1. High-resolution STEM investigation of a multilayer *h*-BN.** (a) Plan-view STEM-MAADF image of the multilayer *h*-BN. The inset shows the Fast Fourier Transform (FFT) pattern for the STEM image. (b-c) Cross-sectional STEM images of *h*-BN multilayer cutting along the [10-10] (Armchair edge) (b) and [11-20] (Zigzag edge) (c) orientations, respectively. (d) Schematic top view of *h*-BN crystal with the crystal lattice parameters. (e) 3D schematic of a *h*-BN bilayer in the AA′ stacking.

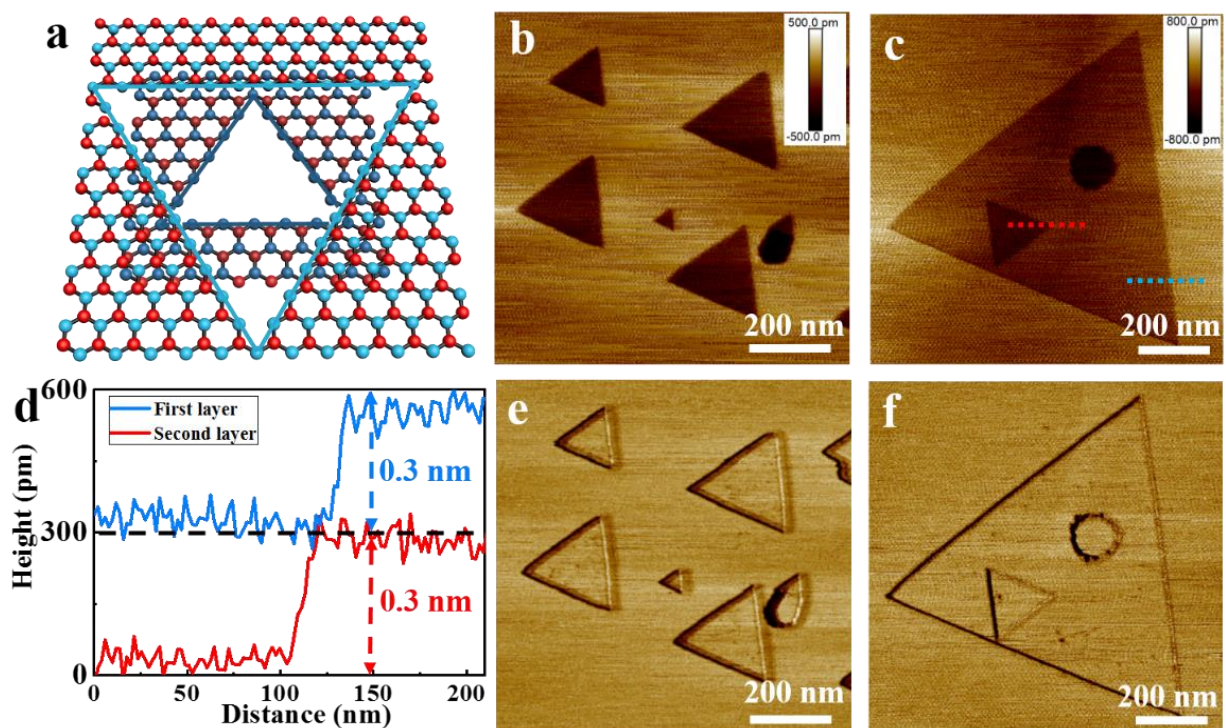

**Figure 2. Anisotropic etching in multilayer *h*-BN.** (a) Schematic of an AA′ stacked *h*-BN bilayer etched with triangle-shaped holes with edges of the same termination. (b) Atomic force microscopy (AFM) height image of multilayer *h*-BN where only the top *h*-BN layer is etched. (c) Height image on multilayer *h*-BN where the second layer of *h*-BN was also etched. (d) Height profiles along the dotted lines in (c). (e-f) Corresponding AFM friction images of (b-c).

The hexagonal pattern shown in the inset of **Figure 1a** was obtained by Fast Fourier Transform (FFT) and also consistent well with the AA′ stacking order of multilayer *h*-BN.

Cross-sectional TEM/STEM measurements were also carried out to confirm the stacking order of the *h*-BN multilayer. Two different slabs with the orientations along [10-10] (armchair edge) and [11-20] (zigzag edge) (Figure S1) were obtained by focused ion beam (FIB) cutting. **Figure 1b** shows the cross-sectional STEM-MAADF image of the multilayer *h*-BN cut along [10-10]. The simulated diffraction pattern of *h*-BN is also shown in the Supporting Information Figure S4, and is in good agreement with the experimental results. In this figure, each bright spot is a combination of two atomic columns, as observed from the atomic model presented in the inset of **Figure 1b**, because the spacing between the nearest-neighbor boron and nitrogen atoms is only 0.72 Å, which is below the STEM resolution. The intensity line profiles of the lattice are shown in the Supporting Information Figure S3. The cross-sectional STEM image of the multilayer *h*-BN cut along the [11-20] direction is shown in **Figure 1c**, and the simulated result is shown in the Supporting Information Figure S5. Unlike in Figure 1b, in this STEM image (**Figure 1c**), each bright spot corresponds to an atomic column.

In addition to TEM/STEM measurements, the etching of *h*-BN at high temperature was also carried out to investigate the stacking order. The multilayer *h*-BN samples obtained by mechanical exfoliation were first annealed in argon/oxygen at 900ºC in order to clean the surface, and then were heated to 1200ºC in pure oxygen at the pressure of 5-6 Pa. As shown in **Figure 2a**, when two adjacent layers of *h*-BN were etched, the triangle-shaped holes in the upper and lower layers always have opposite orientations because of the AA′ stacking order of *h*-BN. **Figure 2b-f** are the experimental atomic force microscopy (AFM) results. As shown in **Figure 2b**, the triangle-shaped holes of the top layer of *h*-BN have the same orientation. Once the second layer is etched, the orientation of these triangle-shaped holes in the second layer will be opposite to the orientation of the holes in the top layer (see **Figure 2c**). The height profiles presented in **Figure 2d** indicate that the triangle-shaped holes are only one layer deep. Thus, the etching experiment is in agreement with the AA′ stacking order in multilayer *h*-BN. According to the kinetic Wulff construction model, the edges of the holes etched in each layer of *h*-BN should be terminated by the edges with the lowest etching rate [34]. *h*-BN is a binary compound with a layered structure, and therefore, the monoatomic layer etching usually shows anisotropic etching to form a triangle-shaped pattern. Thus, the experimental results confirm that the multilayer *h*-BN is found in the AA′ stacking arrangement.

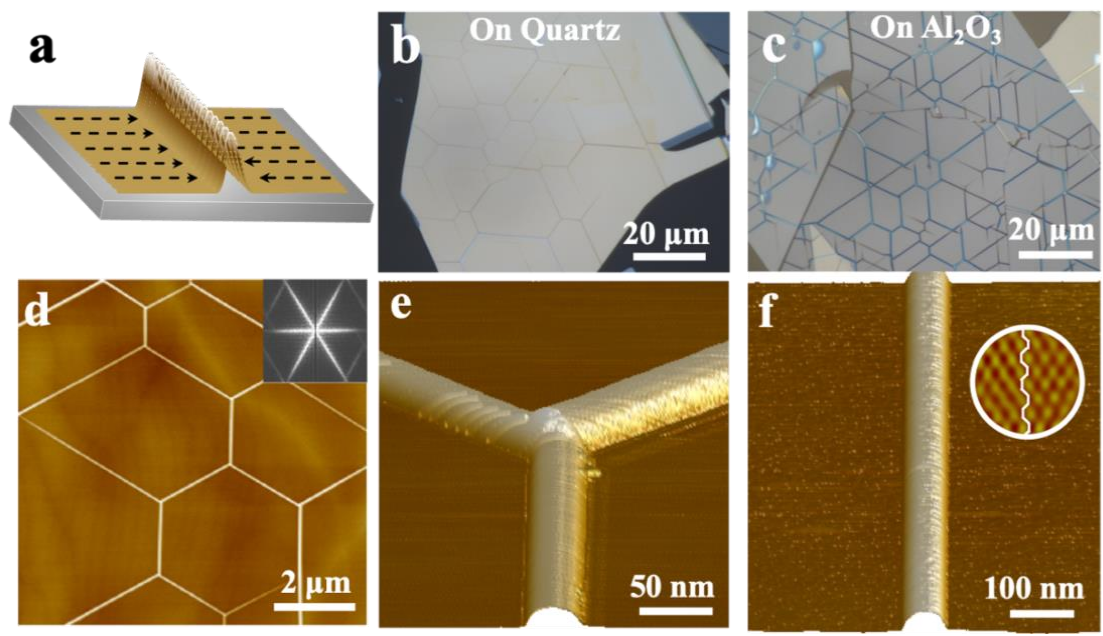

**Figure 3. Crystal-oriented wrinkles on multilayer *h*-BN.** (a) Schematic of *h*-BN wrinkle formation, with the black arrows indicating the direction of the strain. (b) Optical image of wrinkled *h*-BN on quartz after annealing. (c) Optical image of wrinkled *h*-BN on Al₂O₃ after annealing. (d) AFM image of the crystallographically oriented wrinkles on multilayer *h*-BN on quartz. The inset shows the FFT image of this topography. (e) A magnified view of a "Y-type" junction where the wrinkles are oriented by 120° with respect to each other. (f) A magnified AFM image of a straight wrinkle. The circular inset shows a lattice-resolution AFM image of the *h*-BN.

As shown in **Figure 3**, a network of wrinkles is observed on the *h*-BN flakes after high-temperature annealing. **Figures 3b** and **3c** show the optical images of the *h*-BN flakes on the quartz and Al₂O₃ substrates, respectively, after annealing at 1200°C. **Figure 3d** shows the AFM image of annealed *h*-BN on quartz. It is observed from **Figure 3d**, that the wrinkles always exhibit separation angles of ~60°. The FFT image inserted in **Figure 3d** shows that the wrinkles are oriented along the specific crystallographic orientations of *h*-BN. The formation of these wrinkles is mainly due to the mismatch in the thermal expansion coefficient between *h*-BN and the substrate. The in-plane thermal expansion coefficient of *h*-BN is negative at room temperature while it is positive in most materials, including SiO₂ and Al₂O₃. Therefore, *h*-BN on the substrate is subjected to compressive stress after cooling, leading to the formation of the wrinkles (**Figure 3a**). These wrinkles either form "Y-type" junctions or remain as straight dislocations as shown in **Figures 3e** and **3f**. The formation of the wrinkles is related to the threefold symmetry of the *h*-BN lattice. The lattice-resolution AFM image shows that the wrinkle is oriented along the armchair direction of *h*-BN (See inset in **Figure 3f**). It is clear that the wrinkles can be used to easily determine the crystallographic orientation of *h*-BN [9-11].

The wrinkles were observed by STEM, with the beam energy maintained at 60 keV in order to minimize the damage to the lattice. The results are shown in **Figure 4**. Since the two sidewalls of the wrinkle are approximately parallel to the incident direction of the electron beam, the thickness of the wrinkle can be measured directly. **Figures 4a** shows that the thickness of the wrinkle is 5 atomic layers. A magnified view shows that the lattice spacings in the lateral and vertical directions are ~0.22 nm and ~0.34 nm, respectively (**Figures 4b-c,** and Figure S6 in the supporting information). The wrinkle is clearly oriented along armchair orientation..The electron energy loss spectroscopy (EELS) measurement is also carried out to investigate *h*-BN wrinkles (Figure S7). **Figure 4d** shows the cross-sectional STEM bright-field (BF) image of a *h*-BN wrinkle corner that indicates the strain zone is located at a small region as shown in the dashed square (more details are shown in Figure S8).

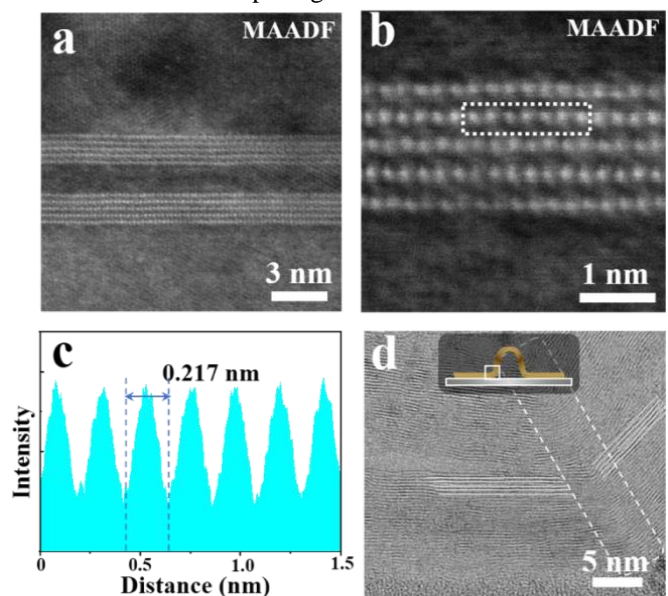

**Figure 4. STEM investigation of crystal-oriented wrinkles on multilayer *h*-BN.** (a) A STEM-MAADF image of a wrinkle on multilayer *h*-BN. (b) Magnified view of (a) showing a sidewall of the *h*-BN wrinkle. (c) Profile of the image intensity along the longitudinal axis of the white dashed frame in (b). (d) Cross-sectional STEM image of a *h*-BN wrinkle corner. The dashed frame inserted shows the strain region.

An examination of **Figure 4a** shows that the inner width of the empty channel in the wrinkle is less than 2 nm, which is on the order of molecular size, enabling the wrinkle to serve as a transport channel for ions or water molecules. The width, height and density of the wrinkles can be controlled by changing the *h*-BN thickness, substrate type and annealing temperature, as described in greater detail in the Supporting Information (Figures S9-11, Table S1). Recent research has shown that small ions and water molecules can be transported through patterned nanocapillaries of 2D materials obtained by lithography patterning and transfer stacking techniques [25]. Compared with those nanocapillaries, a wrinkle-based channel can be easily implemented and does not suffer from the presence of dangling bonds. Therefore, such nanowrinkles may be useful for the development of nanoparticle separation and nanofluidics.

The geomtric analysis and theoretical simulation were carried out to understand the mechanism of wrinkle formation. As shown in Figure S12, the *h*-BN lattice is easy to wrinkle along armchair direction as the density of B-N bonds bent is the lowest. The geometric analysis was shown in the Supporting Information. In the simulation, the energy variations of wrinkling along the zigzag and armchair directions in *h*-BN multilayers were investigated by molecular dynamics (MD) simulations that were performed using the LAMMPS package [35]. The atomic structures of the wrinkles in the AA′ stacked multilayer *h*-BN after relaxation are shown in **Figures 5c** and **d**, respectively. The MD simulations and force field parameters are described in greater detail in the supporting information.

The energy difference $\Delta E$ between the wrinkled and flat states was calculated as $\Delta E = (E_w - E_f)/S$, where $E_w$ is the total energy of the wrinkled *h*-BN multilayer, $E_f$ is the total energy of the flat *h*-BN multilayer, and $S$ is the area of flat *h*-BN. As shown in **Figure 5e**, $\Delta E$ of the wrinkle along the zigzag direction is always larger than that of the wrinkle along the armchair direction. The energy difference $\Delta E_{z\text{-}a}$ between the wrinkle along the zigzag direction and the wrinkle along the armchair directions is given by $\Delta E_{z\text{-}a} = \Delta E_z - \Delta E_a$, where $\Delta E_z$ and $\Delta E_a$ are the energy differences between the wrinkled and flat states for the zigzag and armchair directions, respectively. As shown in **Figure 5f**, it is found that $\Delta E_{z\text{-}a}$ is always positive and $\Delta E_{z\text{-}a}$ increases monotonically with the number of layers *n* for *n* greater than 9. A lower energy difference $\Delta E$ means a smaller energy barrier to wrinkling. Therefore, the wrinkling of a *h*-BN multilayer in the AA′ stacking is energetically favorable along the armchair direction, which is in good agreement with the experimental results.

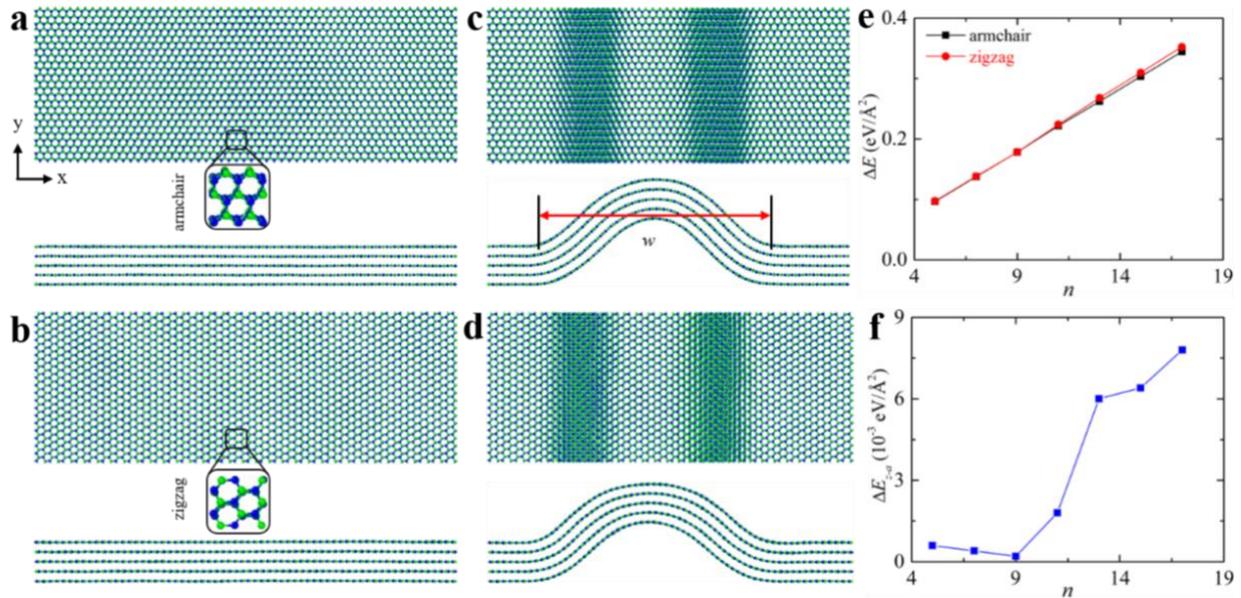

**Figure 5.** Atomic configurations of wrinkled *h*-BN multilayers along the (a) (c) armchair and (b) (d) zigzag directions. The number of layers is set to 5, and the green and blue spheres represent B and N atoms, respectively. The insets in (a) and (b) show schematics of atomic structures of *h*-BN. For both armchair and zigzag edges, the width *w* of the wrinkle is approximately 7 nm. (e) Energy difference $\Delta E$ between the wrinkled and flat *h*-BN multilayers as a function of the layer number. (f) Energy difference $\Delta E_{z\text{-}a}$ between the *h*-BN wrinkles formed along the zigzag and armchair directions. Here, *n* is the number of layers.

## 3. Conclusion

In conclusion, the lattice structure and stacking order in multilayer $h$-BN were investigated by TEM/STEM. The lattice constant and stacking order are in good agreement with the theoretical results reported in the literature. Anisotropic etching of $h$-BN always produces triangle-shaped holes with a depth of atomic monolayer. It was observed that the triangle-shaped holes obtained in the adjacent layers always have opposite orientations. This indicates that the multilayer $h$-BN is in the AA′ stacking. Furthermore, the negative thermal expansion coefficient of the $h$-BN layers often leads to wrinkle formation in heating-cooling cycles. STEM observations reveal that the wrinkles of multilayer $h$-BN are oriented along armchair direction. Further theoretical results show that wrinkling along the armchair direction is energetically favorable. These finding may be helpful for rapid and easy evaluation of the crystallographic orientation of bulk $h$-BN. Additionally, wrinkle formation may play an important role in strain engineering and nanofluidics in the future.

## Acknowledgements

The work was partially supported by the National Key R&D program (Grant No. 2017YFF0206106), the Strategic Priority Research Program of Chinese Academy of Sciences (Grant No. XDB30000000), the National Science Foundation of China (Grant Nos. 51772317, 91964102, 12004406, 11622218, 11972186, 11890674, 51921003), the Science and Technology Commission of Shanghai Municipality (Grant No. 16ZR1442700), the Shanghai Post-doctoral Excellence Program, the China Postdoctoral Science Foundation (Grant Nos. 2019T120366, 2019M651620), the NSF of Jiangsu Province (BK20160037), the Fundamental Research Funds for the Central Universities (NO. NE2019001) of China, and a Project Funded by the Priority Academic Program Development of Jiangsu Higher Education Institutions. K.W. and T.T. acknowledge support from the Elemental Strategy Initiative conducted by the MEXT, Japan and JSPS KAKENHI Grant Numbers JP15K21722. L.C. and H.W. thank Shilong Lv and Tianjiao Xin (Microstructural Characterization Platform in Shanghai Institute of Microsystem and Information Technology, Chinese Academy of Sciences) for FIB and TEM measurement.

## Supplementary information

More inforamiton of the TEM results and the molecular dynamics simulations are available on Supplementntary Information.

## References

[1] Yin J, Yu J, Li X, Li J, Zhou J, Zhang Z and Guo W 2015 Large single-crystal hexagonal boron nitride monolayer domains with controlled morphology and straight merging boundaries *Small* **11** 4497-4502.

[2] Watanabe K, Taniguchi T and Kanda H 2004 Direct-bandgap properties and evidence for ultraviolet lasing of hexagonal boron nitride single crystal *Nat. Mater.* **3** 404-409.

[3] Kim S M *et al.* 2015 Synthesis of large-area multilayer hexagonal boron nitride for high material performance *Nat. Commun.* **6** 8662.

[4] Caldwell J D, Aharonovich I, Cassabois G, Edgar J H, Gil B and Basov D N 2019 Photonics with hexagonal boron nitride *Nat. Rev. Mater.* **4** 552-567.

[5] Pakdel A, Bando Y and Golberg D 2014 Nano boron nitride flatland *Chem. Soc. Rev.* **43** 934-959.

[6] Dean C R *et al.* 2010 Boron nitride substrates for high-quality graphene electronics *Nat. Nanotechnol.* **5** 722-726.

[7] Tang S, Ding G, Xie X, Chen J, Wang C, Ding X, Huang F, Lu W and Jiang M 2012 Nucleation and growth of single crystal graphene on hexagonal boron nitride *Carbon* **50** 329-331.

[8] Tang S *et al.* 2015 Silane-catalysed fast growth of large single-crystalline graphene on hexagonal boron nitride *Nat. Commun.* **6** 6499.

[9] Chen L *et al.* 2017 Oriented graphene nanoribbons embedded in hexagonal boron nitride trenches *Nat. Commun.* **8** 14703.

[10] Chen L *et al.* 2017 Edge control of graphene domains grown on hexagonal boron nitride *Nanoscale* **9** 11475-11479.

[11] Tang S *et al.* 2013 Precisely aligned graphene grown on hexagonal boron nitride by catalyst free chemical vapor deposition *Sci. Rep.* **3** 2666.

[12] Constantinescu G, Kuc A and Heine T 2013 Stacking in bulk and bilayer hexagonal boron nitride *Phys. Rev. Lett.* **111** 036104.

[13] Hao Y, Wang Y, Wang L, Ni Z, Wang Z, Wang R, Koo C K, Shen Z and Thong J T 2010 Probing layer number and stacking order of few-layer graphene by Raman spectroscopy *Small* **6** 195-200.

[14] Lui C H, Li Z, Chen Z, Klimov P V, Brus L E and Heinz T F 2011 Imaging stacking order in few-layer graphene *Nano Lett.* **11** 164-169.

[15] Kim D S, Kwon H, Nikitin A Y, Ahn S, Martin-Moreno L, Garcia-Vidal F J, Ryu S, Min H and Kim Z H 2015 Stacking structures of few-layer graphene revealed by phase-sensitive infrared nanoscopy *ACS Nano* **9** 6765-6773.

[16] Lipp A, Schwetz K A and Hunold K 1989 Hexagonal boron nitride: Fabrication, properties and applications *J. Eur. Ceram. Soc.* **5** 3-9.

[17] He L *et al.* 2019 Isolating hydrogen in hexagonal boron nitride bubbles by a plasma treatment *Nat. Commun.* **10** 2815.

[18] Lim H, Jung J, Ruoff R S and Kim Y 2015 Structurally driven one-dimensional electron confinement in sub-5-nm graphene nanowrinkles *Nat. Commun.* **6** 8601.

[19] Lyu B *et al.* 2019 Phonon polariton-assisted infrared nanoimaging of local strain in hexagonal boron nitride *Nano Lett.* **19** 1982-1989.

[20] Jiang Y, Mao J, Duan J, Lai X, Watanabe K, Taniguchi T and Andrei E Y 2017 Visualizing strain-induced pseudomagnetic fields in graphene through an hBN magnifying glass *Nano Lett.* **17** 2839-2843.

[21] Bao W, Miao F, Chen Z, Zhang H, Jang W, Dames C and Lau C N 2009 Controlled ripple texturing of suspended

graphene and ultrathin graphite membranes *Nat. Nanotechnol.* **4** 562-566.

[22] Chen C C, Bao W, Theiss J, Dames C, Lau C N and Cronin S B 2009 Raman spectroscopy of ripple formation in suspended graphene *Nano Lett.* **9** 4172-4176.

[23] Levy N, Burke S A, Meaker K L, Panlasigui M, Zettl A, Guinea F, Castro Neto A H and Crommie M F 2010 Strain-induced pseudo-magnetic fields greater than 300 tesla in graphene nanobubbles *Science* **329** 544-547.

[24] Meng L *et al.* 2020 Wrinkle networks in exfoliated multilayer graphene and other layered materials *Carbon* **156** 24-30.

[25] Esfandiar A, Radha B, Wang F C, Yang Q, Hu S, Garaj S, Nair R R, Geim A K and Gopinadhan K 2017 Size effect in ion transport through angstrom-scale slits *Science* **358** 511-513.

[26] Gopinadhan K, Hu S, Esfandiar A, Lozada-Hidalgo M, Wang F C, Yang Q, Tyurnina A V, Keerthi A, Radha B and Geim A K 2019 Complete steric exclusion of ions and proton transport through confined monolayer water *Science* **363** 145-148.

[27] Fumagalli L *et al.* 2018 Anomalously low dielectric constant of confined water *Science* **360** 1339-1342.

[28] Cai Q, Scullion D, Gan W, Falin A, Zhang S, Watanabe K, Taniguchi T, Chen Y, Santos E J G and Li L H 2019 High thermal conductivity of high-quality monolayer boron nitride and its thermal expansion *Sci. Adv.* **5** eaav0129.

[29] Yates B, Overy M J and Pirgon O 1975 The anisotropic thermal expansion of boron nitride *Philos. Mag.* **32** 847-857.

[30] Oliveira C K *et al.* 2015 Crystal-oriented wrinkles with origami-type junctions in few-layer hexagonal boron nitride *Nano Res.* **8** 1680-1688.

[31] Shi Z *et al.* 2020 Vapor-liquid-solid growth of large-area multilayer hexagonal boron nitride on dielectric substrates *Nat. Commun.* **11** 849.

[32] Meyer J C, Chuvilin A, Algara-Siller G, Biskupek J and Kaiser U 2009 Selective sputtering and atomic resolution imaging of atomically thin boron nitride membranes *Nano Lett.* **9** 2683-2689.

[33] Alem N, Erni R, Kisielowski C, Rossell M D, Gannett W and Zettl A 2009 Atomically thin hexagonal boron nitride probed by ultrahigh-resolution transmission electron microscopy *Phys. Rev. B* **80** 155425.

[34] Ma T, Ren W, Zhang X, Liu Z, Gao Y, Yin L C, Ma X L, Ding F and Cheng H M 2013 Edge-control growth and kinetics of single-crytal graphene domains by chemical vapor deposition *Proc. Natl. Acad. Sci. U. S. A.* **110** 20386-20391.

[35] Plimpton S 1995 Fast parallel algorithms for short-range molecular dynamics *J. Comput. Phys.* **117** 1-19.